\documentclass[a4paper,11pt]{article}
\usepackage{pos}

\usepackage{graphicx}
\usepackage{dcolumn}
\usepackage{bm,graphicx,hyperref,subfigure}
\usepackage{float,amsfonts,graphicx}
\usepackage{color}
\usepackage{tikz}
\usepackage{adjustbox}
\usetikzlibrary{decorations.pathreplacing}
\usetikzlibrary{arrows.meta}
\relax

\DeclareMathOperator{\Tr}{\mathrm{Tr}}

\DeclareMathOperator{\dd}{{\rm{d}}}

\newcommand{\redchisq}{\chi^2_{\tiny\mbox{red}}}
\newcommand{\lambdai}{\lambda_{\mbox{\tiny{i}}}}
\newcommand{\lambdaf}{\lambda_{\mbox{\tiny{f}}}}
\newcommand{\betac}{\beta_{\mbox{\tiny{c}}}}

\newcommand{\eq}[1]{\begin{align}\label{#1}}
\newcommand{\en}{\end{align}}
\newcommand{\eqar}[1]{\begin{align}\label{#1}}
\newcommand{\enar}{\end{align}}

\title{Entanglement entropy from non-equilibrium lattice simulations}

\author*[a]{Andrea Bulgarelli}
\author[a]{Marco Panero}

\affiliation[a]{Physics Department, University of Turin \& INFN, Turin unit\\Via Pietro Giuria 1, I-10125 Turin, Italy}

\emailAdd{andrea.bulgarelli@unito.it}
\emailAdd{marco.panero@unito.it}

\abstract{Entanglement entropy encodes important features of strongly interacting quantum many-body systems and gauge theories, but its analytical study is still limited to systems with high level of symmetry. This motivates the search for efficient techniques to investigate this quantity numerically, through Monte Carlo calculations on the lattice. In this contribution, we discuss the computation of the entropic c-function by means of an algorithm based on Jarzynski's equality, which is an exact theorem from non-equilibrium statistical mechanics. After presenting benchmark results for the Ising model in two dimensions, where our algorithm successfully reproduces the analytical predictions from conformal field theory, we discuss its generalization to the three-dimensional Ising model, for which we were able to extract universal terms beyond the area law. Finally we point out some future generalizations of this calculation.}

\FullConference{The 40th International Symposium on Lattice Field Theory (Lattice 2023)\\
July 31st - August 4th, 2023\\
Fermi National Accelerator Laboratory\\}

\begin{document}
\maketitle

\section{Introduction}
Entanglement is a key property of quantum systems, with far-reaching implications in condensed matter~\cite{Laflorencie:2015eck}, in high-energy physics~\cite{Klebanov:2007ws} and in quantum computing~\cite{nielsen2000quantum}. For a system made up of two components $A$ and $B$, entanglement can be quantified by the entanglement entropy~\cite{Bennett:1995tk}, i.e., by the von~Neumann entropy of $\rho_A = \Tr_B \rho$, the reduced density matrix associated with $A$:
\begin{align}
S(A) = -\Tr(\rho_A\ln\rho_A).
\end{align}
$S(A)$ can be obtained as the $n \to 1$ limit of $S_n(A) = \frac{1}{1-n}\ln\Tr\rho_A^{n}$, the R\'enyi entropy of order $n$. While the leading term in the entanglement entropy is proportional to the area $|\partial A|$ of the entangling surface separating $A$ and $B$, and is ultraviolet-divergent, subleading contributions to $S(A)$ are finite, and encode important physical information. Considering a system defined in a $(d+1)$-dimensional torus, and letting $A$ be a slab of length $l$ in the direction separating $A$ and $B$, and maximally extended in the other directions, the subleading contributions to $S(A)$ can be studied in terms of entropic (R\'enyi) c-functions~\cite{Casini:2004bw}
\begin{align}
\label{definition_entropic_c-function}
C_n(l) = \frac{l^{d}}{|\partial A|}\frac{\partial S_n}{\partial l},
\end{align}
which ``count the degrees of freedom'' of the theory~\cite{Zamolodchikov:1986gt}.

While the analytical computation of entropic c-functions can be carried out only for highly symmetric systems, also their numerical estimate is difficult, due to the non-local nature of the observable, to the high computational costs to sample the space of configurations, and to the limited scalability to systems defined in more than $1+1$ dimensions~\cite{Buividovich:2008kq,Caraglio:2008pk,Itou:2015cyu,Rabenstein:2018bri,Rindlisbacher:2022bhe,Jokela:2023yun}. In this contribution we present a new method to tackle these challenges and to evaluate the entropic c-function by means of non-equilibrium Monte~Carlo simulations, that we recently applied to the Ising model in two and in three dimensions~\cite{Bulgarelli:2023ofi}. In the following, we use natural units, denoting the coupling between spins in units of the temperature as $\beta$ and the linear extent of the lattice in the spatial directions as $L$.

\section{Simulation algorithm}
\begin{figure*}[t]
\centering
\begin{tikzpicture}[scale=0.7, every node/.style={scale=0.7}]
\draw[step = 1cm, gray, thin] (-.9+ 10, -.9) grid (5.9+ 10, 5.9);
\foreach \x in {10,...,15}{
  \foreach \y in {0,...,5}{
    \fill[color=black] (\x, \y) circle (.1);
  }
}
\draw[cyan, ultra thick] (2.5+ 10,-.9)--(2.5+ 10,5.9);
\draw (4.5+10, 1.5)node{$A$};
\draw (2.7+10, 3.5)node{$\partial A$};
\draw (0.5+10, 1.5)node{$B$};

\foreach \y in {0,...,5}{
  \fill[color=orange] (2+10, \y) circle (.1);
}

\draw[decorate,decoration={brace,amplitude=5}](-.9,6.2)--(1.4,6.2);
\draw[decorate,decoration={brace,amplitude=5}](1.6,6.2)--(5.9,6.2);
\draw[decorate,decoration={brace,amplitude=5}](-1.2,-.9)--(-1.2,5.9);
\draw (.2,6.8)node{$L-l-a$};
\draw (3.7,6.8)node{$l+a$};
\draw (-1.8,2.5)node{$L$};
\draw[ultra thick,-Stealth, red] (6.5,2.5)--(8.5,2.5);
\draw[step = 1cm, gray, thin] (-.9, -.9) grid (5.9, 5.9);
\foreach \x in {0,...,5}{
  \foreach \y in {0,...,5}{
    \fill[color=black] (\x, \y) circle (.1);
  }
}
\draw[cyan, ultra thick] (1.5,-.9)--(1.5,5.9);
\draw (4.5, 1.5)node{$A$};
\draw (1.7, 3.5)node{$\partial A$};
\draw (0.5, 1.5)node{$B$};

\foreach \y in {0,...,5}{
  \fill[color=orange] (2, \y) circle (.1);
}

\draw[decorate,decoration={brace,amplitude=5}](-.9+10,6.2)--(2.4+10,6.2);
\draw[decorate,decoration={brace,amplitude=5}](2.6+10,6.2)--(5.9+10,6.2);
\draw (.7+10,6.8)node{$L-l$};
\draw (4.2+10,6.8)node{$l$};
\end{tikzpicture}
\caption{Sketch of the algorithm to compute the entropic c-function. The figure shows a spatial slice of a single replica in a three-dimensional system; sites belonging to $A$ are connected to another replica through links in the Euclidean-time direction, while those belonging to $B$ are connected to the same replica. The ratio of partition functions in eq.~(\ref{entropic_c-function_in_code}) is computed from the exponential average of the work done on the system by varying the couplings of yellow sites from the initial geometry, where they are connected to another replica, to the one at the end of the non-equilibrium evolution, where they are connected to the same replica.}
\label{fig:higher_dimensions_c-function_algorithm}
\end{figure*}
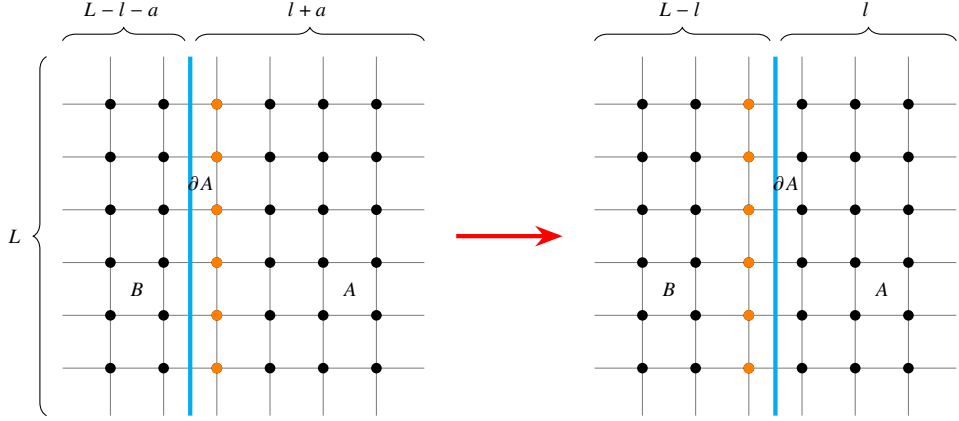

Our Monte~Carlo algorithm combines the replica trick~\cite{Calabrese:2004eu} and Jarzynski's theorem~\cite{Jarzynski:1996oqb} to evaluate the entropic c-function. Given a statistical system whose Hamiltonian depends on a set of parameters $\lambda$, Jarzynski's equality relates the partition functions corresponding to different values $\lambdai$ and $\lambdaf$ to the work $W$ (in units of the temperature $T$) that is done on the system, when it is driven out of equilibrium by varying $\lambda$ from $\lambdai$ to $\lambdaf$ as a function of time, $\lambda(t)$:
\begin{align}
\label{Jarzynski_theorem}
\frac{Z_{\lambdaf}}{Z_{\lambdai}} = \overline{ \exp \left( -W/T \right) }.
\end{align}
The bar on the right-hand side denotes the average over non-equilibrium evolutions of the system for a fixed but arbitrary $\lambda(t)$. Jarzynski's equality has been recently applied in a number of lattice simulations~\cite{Caselle:2016wsw, Alba:2016bcp, Caselle:2018kap, DEmidio:2019usm, Francesconi:2020fgi, Zhao:2021njg} and has a direct connection to machine-learning techniques~\cite{Caselle:2022acb}. In ref.~\cite{Bulgarelli:2023ofi}, we proposed to use eq.~\eqref{Jarzynski_theorem} to compute the entropic R\'enyi c-function by means of non-equilibrium simulations of a system in which we directly implemented the replica geometry. The trace of the $n$-th power of the reduced density matrix appearing in the definition of the R\'enyi entropy is then expressed as the ratio of the partition function of the full replica space (denoted as $Z_n$) to the $n$-th power of the partition function of the system, and, using eq.~\eqref{definition_entropic_c-function}, the R\'enyi entropic c-function can be written as
\begin{align}
\label{entropic_c-function_in_code}
C_n(l) = \frac{l^{2}}{|\partial A|}\frac{1}{n-1}\lim_{\epsilon\to 0}\frac{1}{\epsilon}\ln\frac{Z_n(l)}{Z_n(l+\epsilon)}.
\end{align}
We set $\epsilon$ to the lattice spacing $a$ and evaluated the ratio of partition functions on the right-hand side of eq.~\eqref{entropic_c-function_in_code} using eq.~\eqref{Jarzynski_theorem}, computing the work done when one of the two connected components of the entangling surface is shifted by $a$ in a non-equilibrium evolution of the system. The shift of the entangling surface was performed by reducing the coupling of spins near the entangling surface to spins in another replica linearly as a function of Monte~Carlo time (while increasing their coupling to spins within the same replica at the same time), see fig.~\ref{fig:higher_dimensions_c-function_algorithm}. We implemented this algorithm in CUDA, using cluster updates~\cite{Swendsen:1987ce} as described in ref.~\cite{Komura:2012gbs}, and run the code on computers equipped with graphics processing units. Our algorithm works in arbitrary integer dimension $D=d+1$ and can be easily generalized to any spin model with nearest-neighbor interactions.

\section{Results}
\begin{figure*}[t]
\centerline{\includegraphics[height=0.18\textheight]{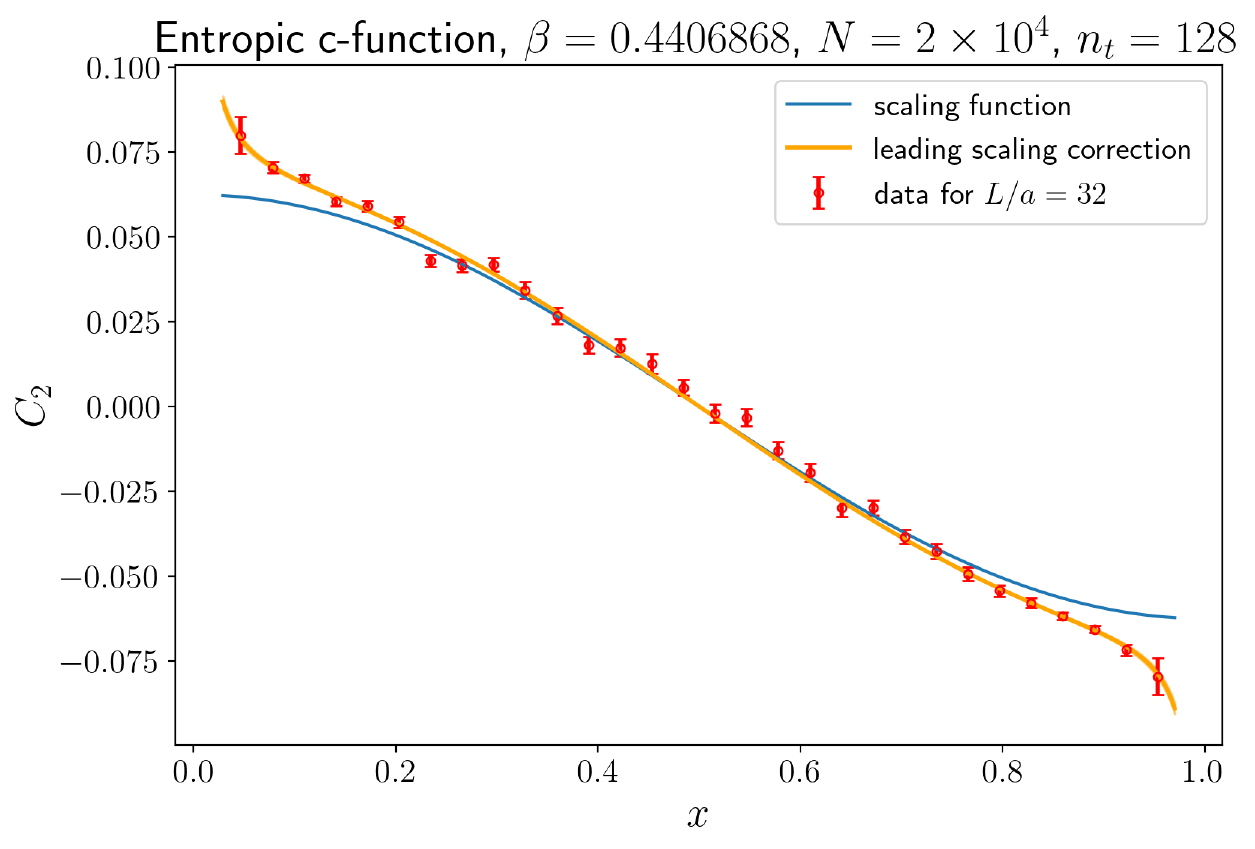} \hfill \includegraphics[height=0.18\textheight]{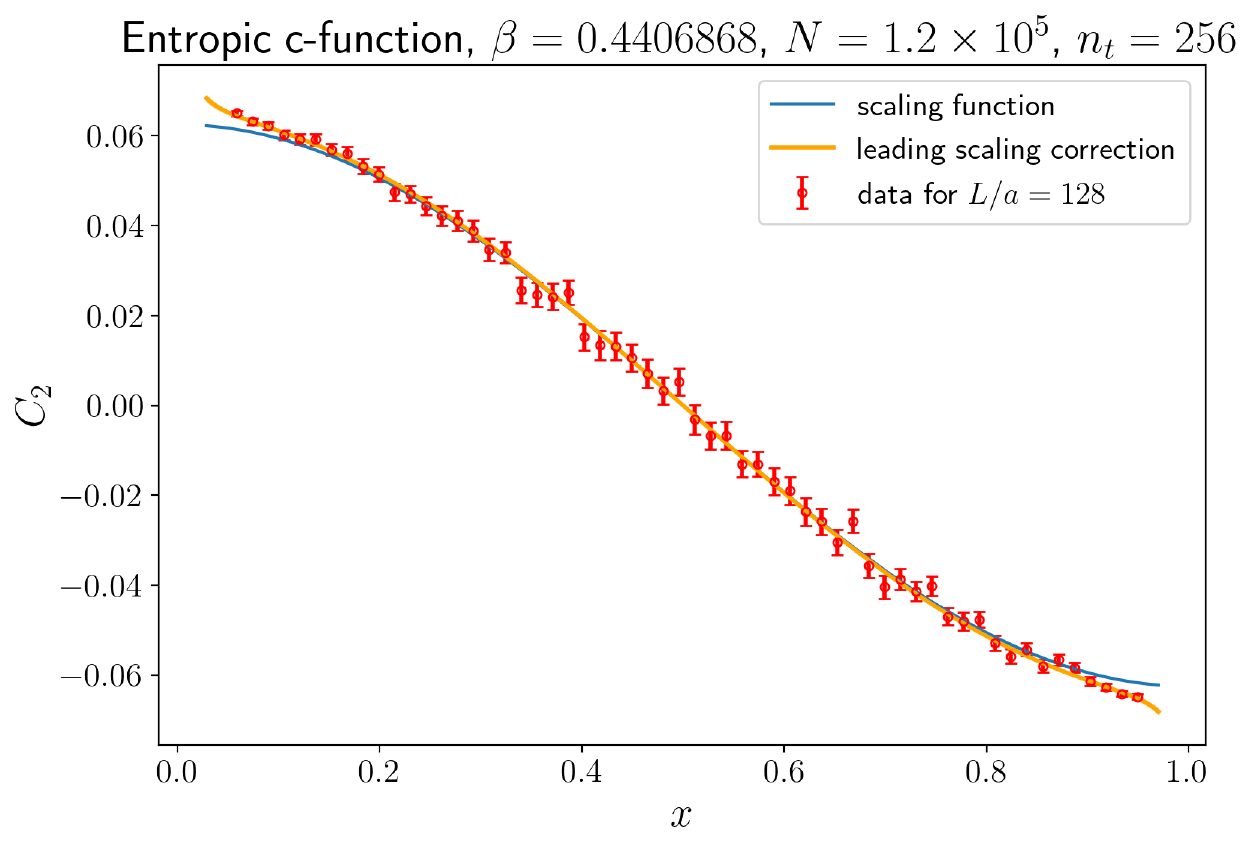} \hfill \includegraphics[height=0.18\textheight]{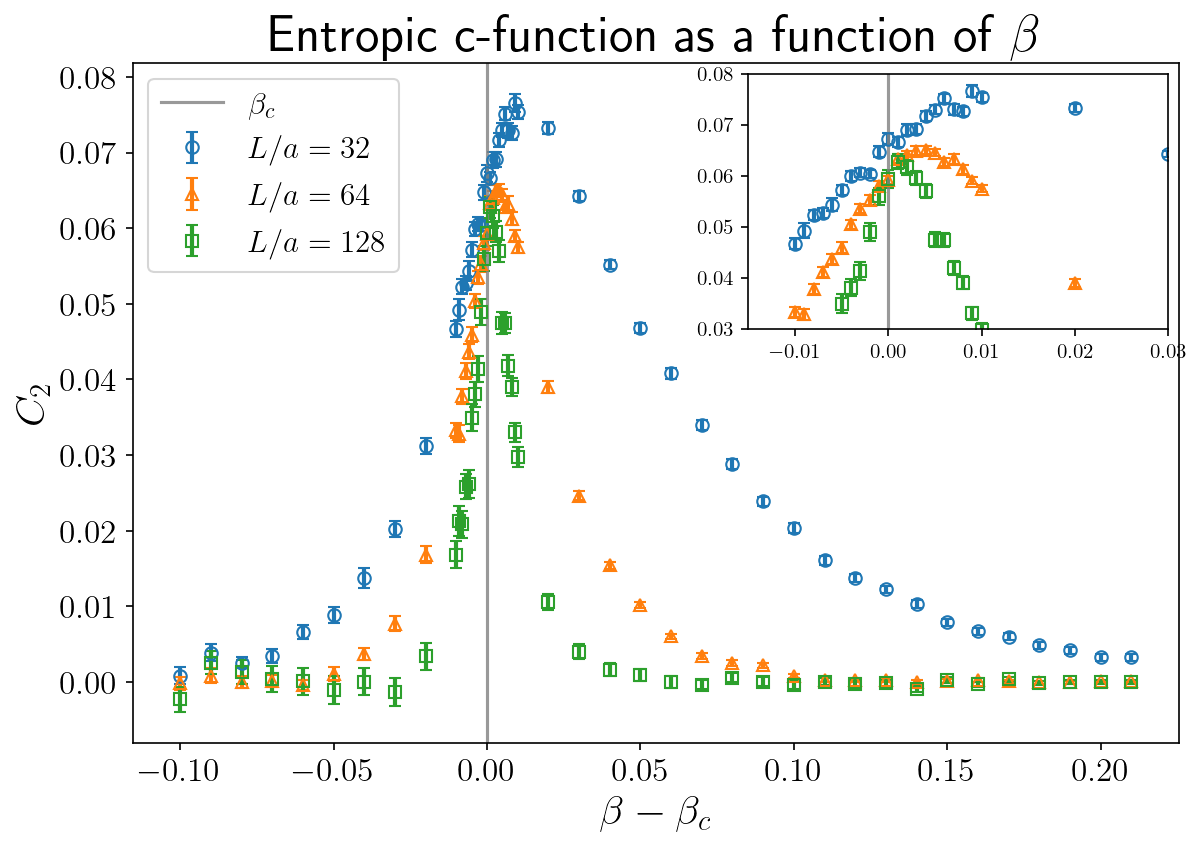}}
\caption{Results for the Ising model in two dimensions: the first two panels show the entropic c-function at the critical point, on a lattice with $L/a=32$ and $L/a = 128$, respectively, and their comparison with analytical predictions from CFT. The last panel shows $C_2$ as a function of $\beta-\betac$ for $L/a = 32$, $64$ and $128$.} \label{fig:results_2d}
\end{figure*}

We first present the results of our simulations for the two-dimensional Ising model. At the critical point of the model (and trading the length of the cut for the arc-length, $l\to\frac{L}{\pi}\sin\left(\frac{\pi l}{L}\right)$, which enforces the symmetry under interchanging $A$ and $B$), conformal field theory (CFT) predicts the second-order entropic c-function to be~\cite{Calabrese:2004eu,Cardy:2010zs}
\begin{align}
    C_2(x)= \frac{c}{8}\cos\left(\pi x\right) + \frac{k}{2L}\cot\left(\pi x\right),
    \label{analytical_c_function_2d}
\end{align}
where $c=1/2$ is the central charge, $x=l/L$, and the second summand on the right-hand side is a finite-size correction. In the first two plots of fig.~\ref{fig:results_2d} we compare our data with the leading scaling correction alone and with eq.~\eqref{analytical_c_function_2d}: for the largest spatial size that we simulated, $L/a = 128$, the scaling function perfectly describes our data ($\redchisq = 1.08$), while for smaller lattices finite-size corrections are non-negligible, but they are captured by the subleading term of eq.~\eqref{analytical_c_function_2d}. The second plot of fig.~\ref{fig:results_2d} shows the behavior of the entropic c-function away from the critical point: as expected, up to finite-size corrections, the function has a maximum corresponding to the critical point and decreases monotonically away from criticality.

\begin{figure*}[t]
\centerline{\includegraphics[height=0.18\textheight]{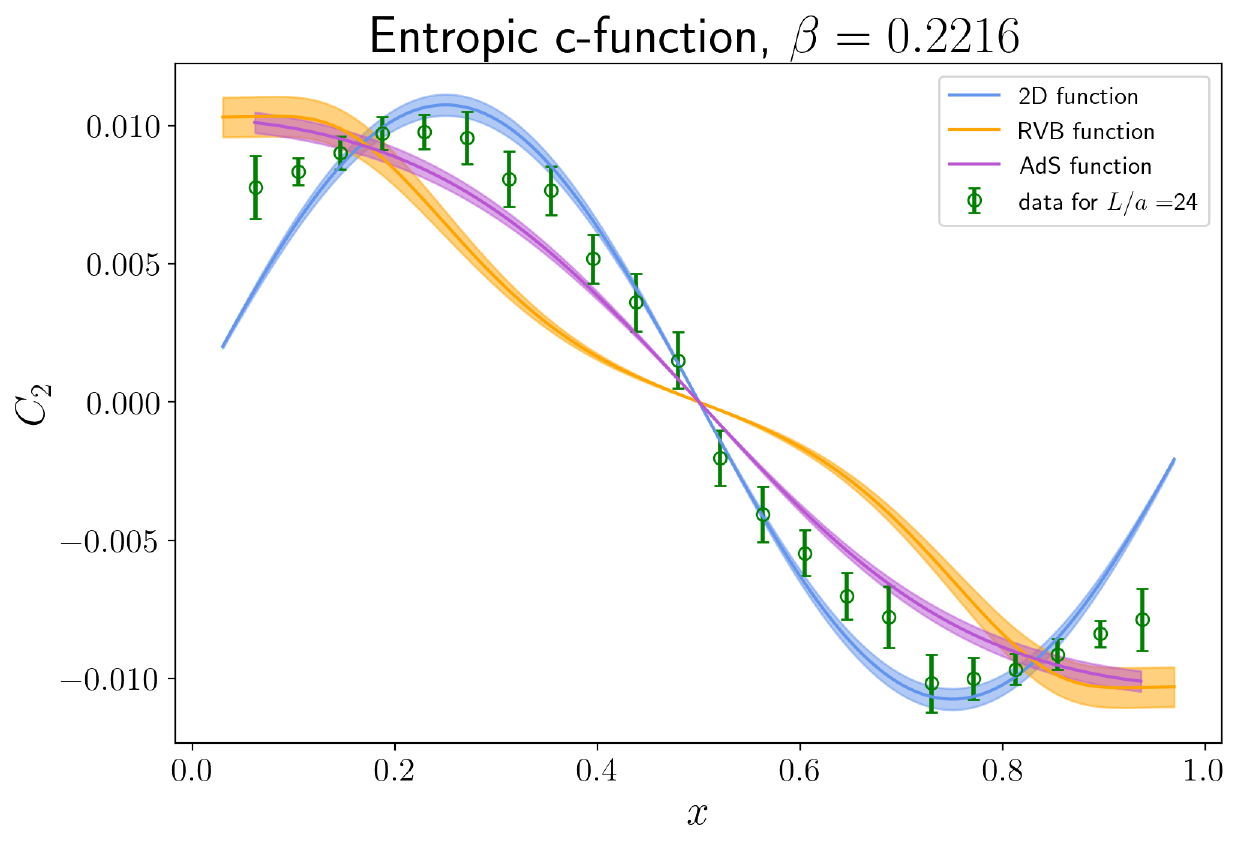} \includegraphics[height=0.18\textheight]{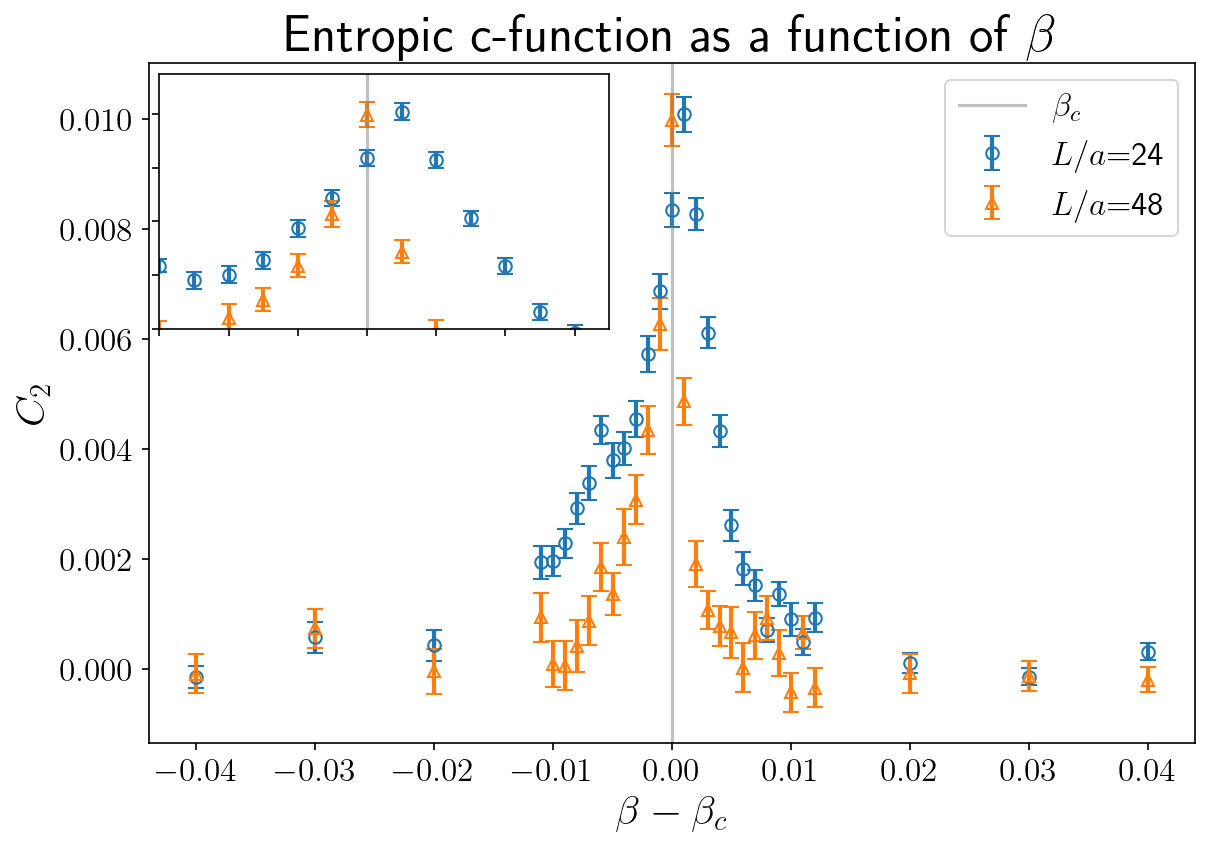}}
\caption{Results for the Ising model in three dimensions: the first panel shows our results for the entropic c-function at the critical point on a lattice with $L/a=24$ and the predictions from the models discussed in the main text. The second panel shows $C_2$ as a function of $\beta-\betac$ for $L/a = 24$ and $48$.} \label{fig:results_3d}
\end{figure*}

For the three-dimensional Ising model we present the results of our simulations at $\beta=0.2216$, which is close to the critical point, $\betac=0.221654626(5)$~\cite{Ferrenberg:2018zst}. In the first plot in fig.~\ref{fig:results_3d}, we compare our numerical data for the entropic c-function with three different models,
\begin{align}
g_{\mbox{\tiny{2D}}}(x;c,k) &= c\ln(\sin(\pi x)) + k, \label{fit_entropy_2d_analogue_3d} \\
g_{\mbox{\tiny{RVB}}}(x;c,k) &= -2 c \ln\left\{ \frac{\eta(\tau)^2}{\theta_3(2\tau)\theta_3(\tau/2)} \frac{\theta_3(2x\tau)\theta_3(2(1-x)\tau)}{\eta(2x\tau)\eta(2(1-x)\tau)}\right\} + k, \label{fit_entropy_RVB_3d} \\
g_{\mbox{\tiny{AdS}}}(x;c,k) &= c \chi(x)^{-\frac{1}{3}}\left\{ \int_0^1\frac{\dd y}{y^2}\left( \frac{1}{\sqrt{P(\chi(x),y)}} - 1 \right) - 1 \right\} + k, \label{fit_entropy_AdS_3d}
\end{align}
where in eq.~\eqref{fit_entropy_RVB_3d} $\tau$ is the modular parameter of the spatial cross-section of the system ($\tau = i$ in our simulations), while in eq.~\eqref{fit_entropy_AdS_3d} $P(\chi,y)=1-\chi y^3 - (1-\chi)y^4$ and $x(\chi)$ is defined as $x(\chi) = \frac{3}{2\pi}\chi^{\frac{1}{3}}(1-\chi)^{\frac{1}{2}}\int_0^1\frac{\dd y y^2}{(1-\chi y^3)}\frac{1}{\sqrt{P(\chi,y)}}$. The function in eq.~\eqref{fit_entropy_2d_analogue_3d} is the one that can be derived in $D=2$, while the one in eq.~\eqref{fit_entropy_RVB_3d} was proposed in a study of resonance-valence-bond (RVB) dimers~\cite{Stephan:2013eig}, a model in the quantum Lifshitz universality class; finally, the function in eq.~\eqref{fit_entropy_AdS_3d} was derived in ref.~\cite{Chen:2014zea} using the Ryu-Takayanagi formula~\cite{Ryu:2006bv}. The corresponding fit functions for the entropic c-function are
\begin{align}
f_j (x;c) = \frac{\sin^2(\pi x)}{2\pi^2} \frac{\dd}{\dd x} g_j (x;c,k),
\end{align}
for $j=\mbox{2D}$, $\mbox{RVB}$, or $\mbox{AdS}$. The second plot in fig.~\ref{fig:results_3d} shows our results for the entropic c-function as a function of $\beta$, which indicate that also in three dimensions $C_2$ is monotonically decreasing away from the critical point.

\section{Discussion and future prospects}
In this contribution, we discussed the algorithm to compute the entropic c-function by means of non-equilibrium Monte~Carlo simulations that we proposed in ref.~\cite{Bulgarelli:2023ofi}, and the results that we obtained for the Ising model. After a benchmark of our code in two dimensions, we studied the three-dimensional case, where we compared our simulation results close to the critical point with the predictions of models based on a generalization of the predictions that can be derived in two dimensions, on a model in the quantum Lifshitz universality class, and on holographic calculations, with the latter providing the best fit to our data. Our results away from the critical point show that, for the slab geometry that we considered, the entropic c-function is monotonically decreasing when one moves away from the critical point, as expected for a Zamolodchikov c-function. This result is non-trivial, as the existing proofs of monotonicity of entropic c-functions in three dimensions depend on the shape of the entangling surface~\cite{Casini:2012ei} and have not been formulated for the geometry that we considered. The present work could be generalized in various ways. The algorithm can be readily adapted to any spin model (with discrete or continuous degrees of freedom) with local interactions, on any type of lattice and in any dimension. Similarly, it could be generalized to different entangling surfaces. We leave these possible research directions for future work.

\section*{Acknowledgements}
This work was partially supported by the Spoke 1 ``FutureHPC \& BigData'' of the Italian Research Center on High-Performance Computing, Big Data and Quantum Computing (ICSC) funded by MUR (M4C2-19) -- Next Generation EU (NGEU), by the Italian PRIN ``Progetti di Ricerca di Rilevante Interesse Nazionale -- Bando 2022'', prot. 2022TJFCYB, and by the ``Simons Collaboration on Confinement and QCD Strings'' funded by the Simons Foundation. The simulations were run on CINECA computers. We acknowledge support from the SFT Scientific Initiative of the Italian Nuclear Physics Institute (INFN).


\end{document}